\documentclass[a4paper,fleqn,usenatbib,letters]{mnras}

\usepackage{graphicx,color}	
\usepackage{graphicx}	
\usepackage{amsmath}	
\usepackage{amssymb}	

\definecolor{mygray}{gray}{0.7}

\title[Imprints of the super-Eddington accretion]{Imprints of the super-Eddington accretion on the quasar clustering}

\author[T. Oogi et al.]{
Taira Oogi,$^{1,2}$\thanks{E-mail: taira.oogi@ipmu.jp}
Motohiro Enoki,$^{3}$
Tomoaki Ishiyama,$^{4}$
Masakazu A. R. Kobayashi,$^{5}$
\newauthor{ Ryu Makiya,$^{1,6}$
Masahiro Nagashima,$^{2}$
Takashi Okamoto$^{7}$ and Hikari Shirakata$^{7}$}
\\
$^{1}$Kavli Institute for the Physics and Mathematics of the Universe (WPI), The University of Tokyo Institutes for Advanced Study, \\ \ The University of Tokyo, Kashiwa, Chiba 277-8583, Japan\\
$^{2}$Faculty of Education, Bunkyo University, 3337 Minami-Ogishima, Koshigaya-shi, Saitama 343-8511, Japan\\
$^{3}$Faculty of Business Administration, Tokyo Keizai University, 1-7-34, Minami-cho, Kokubunji-shi, Tokyo 185-8502, Japan\\
$^{4}$Institute of Management and Infomation Technologies, Chiba University, 1-33, Yayoi-cho, Inage-ku, Chiba 263-8522, Japan\\
$^{5}$Faculty of Natural Sciences, National Institute of Technology, Kure College, 2-2-11, Agaminami, Kure, Hiroshima 737-8506, Japan\\
$^{6}$Max-Planck-Institut fur Astrophysik, Karl-Schwarzschild Str. 1, D-85741 Garching, Germany\\
$^{7}$Department of Cosmosciences, Graduates School of Science, Hokakido University, N10 W8, Kitaku, Sapporo 060-0810, Japan\\
}

\date{Accepted XXX. Received YYY; in original form ZZZ}

\pubyear{2017}

\begin{document}
\label{firstpage}
\pagerange{\pageref{firstpage}--\pageref{lastpage}}
\maketitle

\begin{abstract}
Super-Eddington mass accretion has been suggested as an efficient mechanism to grow supermassive black holes (SMBHs).
We investigate the imprint left by the radiative efficiency of the super-Eddington accretion process on the clustering of quasars using a new semi-analytic model of galaxy and quasar formation based on large-volume cosmological $N$-body simulations.
Our model includes a simple model for the radiative efficiency of a quasar, which imitates the effect of photon trapping for a high mass accretion rate.
We find that the model of radiative efficiency affects the relation between the quasar luminosity and the quasar host halo mass.
The quasar host halo mass has only weak dependence on quasar luminosity when there is no upper limit for quasar luminosity.
On the other hand, it has significant dependence on quasar luminosity when the quasar luminosity is limited by its Eddington luminosity.
In the latter case, the quasar bias also depends on the quasar luminosity, and the quasar bias of bright quasars is in agreement with observations.
Our results suggest that the quasar clustering studies can provide a constraint on the accretion disc model.
\end{abstract}




\section{Introduction}

Some observations \citep{2011Natur.474..616M, 2015Natur.518..512W} have found optically bright quasars even at high redshift, $z\sim7$.
These quasars are powered by mass accretion on to supermassive black holes (SMBHs).
The observations suggest that SMBHs with mass $M_{\mathrm{BH}}\sim10^9 M_{\odot}$ already exist at such high redshift.
While some physical processes through which these SMBHs form have been proposed (see \citealt{2010A&ARv..18..279V} and references therein), 
how the SMBHs form during less than 1 Gyr after the Big Bang is still debated.

Super-Eddington mass accretion, for which the mass accretion rate exceeds the Eddington rate
\begin{equation}
\dot{M}_{\mathrm{Edd}} \equiv L_{\mathrm{Edd}}/c^2,
\label{eq:edd}
\end{equation}
has been suggested as an efficient mechanism to grow SMBHs  \citep*[e.g.][]{2005ApJ...633..624V, 2014ApJ...784L..38M, 2014Sci...345.1330A, 2015ApJ...804..148V}, where $L_{\mathrm{Edd}}$ is the Eddington luminosity $L_{\mathrm{Edd}} \equiv 4\pi G M_{\mathrm{BH}} m_{p} c / \sigma_{\scalebox{0.5}{$T$}}$, $m_{p}$ is the proton mass, and $\sigma_{\scalebox{0.5}{$T$}}$ is the Thomson scattering cross-section.
During the super-Eddington accretion, the photons produced in the accretion flow are advected inwards by the optically thick flow and cannot be radiated away \citep[e.g.][]{1978MNRAS.184...53B}.
This type of the accretion flow is described by a well known solution called slim disc \citep[e.g.][]{1988ApJ...332..646A, 2000PASJ...52..133W, 2000PASJ...52..499M}.
The structure of the slim disc becomes geometrically thick.
\citet*{2015MNRAS.452.1922P} show analytically and numerically that in this condition the outflow by radiative feedback has a negligible role, and the BH can accrete 80--100 per cent of the gas mass of the host halo.
Although there are some observations suggesting that some active galactic nuclei (AGNs) are accreting at super-Eddington rates (e.g. \citealt{2014MNRAS.438..672N}; \citealt*{2017MNRAS.468.3663J}), it is still uncertain whether the super-Eddington accretion actually occurs or not.

The luminosity of a quasar with the super-Eddington accretion phase would be limited to several times of the Eddington luminosity $L_{\mathrm{Edd}}$ due to
the `photon trapping' \citep[e.g.][]{2005ApJ...628..368O}.
This means that the radiative efficiency is low at a high accretion rate.
This limitation could affect the clustering of quasars, in particular, at higher redshifts, because accretion rates for quasars tend to be higher at such higher redshifts (\citealt{2009MNRAS.396..423B}; \citealt{2014ApJ...794...69E}).
If this is true, future wide field surveys may clarify how mass accretion occurs.
In this paper, we focus on the slim disc solution, although alternative radiatively inefficient models exist, e.g. ZEro-BeRnoulli Accretion (ZEBRA; \citealt{2014ApJ...781...82C}) and the ADiabatic Inflow-Outflow Solution (ADIOS; \citealt{1999MNRAS.303L...1B}).

In the previous study with our semi-analytic model (\citealt{2016MNRAS.456L..30O}; hereafter O16),
we have shown that the model quasars with a given host halo mass have various Eddington ratios, $L/L_{\mathrm{Edd}} = \eta \dot{M}_{\mathrm{BH}}/\dot{M}_{\mathrm{Edd}}$, and various luminosities
due to a variety of elapsed times from the beginning of the quasar activity.
As a consequence of this, the median mass of quasar host haloes depends only weakly on the quasar luminosity.
In O16, we did not take into account the fact that the radiative efficiency depends on the accretion rate
when considering the super-Eddington accretion \citep[e.g.][]{2015MNRAS.452.1922P}.
This effect changes the Eddington ratios and luminosities of model quasars with a given host halo mass.
This suggests that the behavior of the luminosity for the accretion rate in the slim disc affects the estimated host halo mass and the spatial clustering of quasars.

In this Letter, we investigate the imprint left by the radiative efficiency of the super-Eddington accretion process on the clustering of quasars.
For this purpose, we use a new semi-analytic model of galaxy and quasar formation based on high-resolution $N$-body simulations.

The remainder of this Letter is organized as follows.
In Section~\ref{sec:model}, we outline our semi-analytic model and the models of radiative efficiency of SMBHs.
Section~\ref{sec:result} gives the mass distribution of the quasar host haloes and the quasar bias for our models.
In Section~\ref{sec:discussion}, we discuss the BH growth at high redshift via the super-Eddington accretion, and summarize our results.

\section{Model}
\label{sec:model}

\begin{figure}
	\centering
	\includegraphics[width=0.9\columnwidth]{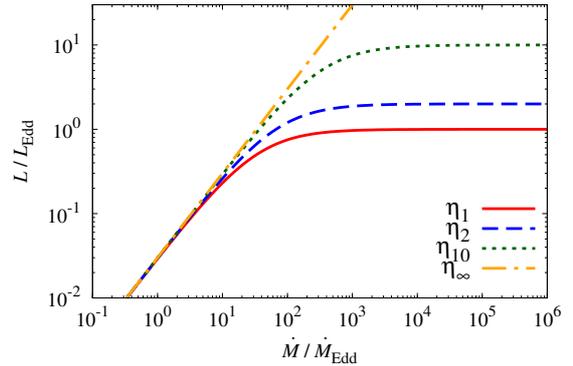}
	\caption{The Eddington ratios as a function of the accretion rate in our models.}
	\label{fig:slimdisk}
\end{figure}

We use a semi-analytic model, $\nu^2$GC (\citealt{2016PASJ...68...25M}), which is an extension of the Numerical Galaxy Catalog ($\nu$GC) \citep{2005ApJ...634...26N, 2014ApJ...794...69E, 2015MNRAS.450L...6S}.
In this model, we adopt merger trees of dark matter (DM) haloes using state-of-the-art cosmological $N$-body simulations (\citealt{2015PASJ...67...61I}) with the Planck cosmology \citep{2014A&A...571A..16P}.
The simulation used in this Letter contains $4096^3$ dark matter particles within a comoving box of 560~$h^{-1}$~Mpc, and the minimum halo mass is $8.79\times 10^9~h^{-1}~M_{\odot}$ (see \citealt{2015PASJ...67...61I} for details).
This simulation enables us to investigate quasar host haloes with statistical significance.
The model includes all the main physical processes involved in galaxy formation: formation and evolution of DM haloes; radiative gas cooling and disc formation in DM haloes; star formation, supernova feedback and chemical enrichment; galaxy mergers.
Free parameters related to galaxy formation processes are set to the values adopted in \citet{2016PASJ...68...25M}.

Our model also includes SMBH growth and quasar formation.
We assume that major mergers of galaxies trigger starbursts in the nuclear regions and cold gas accretion on to SMBHs.
We define a major merger as that with the mass ratio $M_2 / M_1 \geq0.1$, where $M_1$ and $M_2$ are the baryonic masses of the more and less massive galaxies, respectively.
During major mergers, we assume that a fraction of the cold gas is accreted on the SMBH.
The accreted mass $\Delta M_{\mathrm{BH}}$ is modeled as follows:
\begin{equation}
\Delta M_{\mathrm{BH}} = f_{\mathrm{BH}} \Delta M_{\mathrm{star,burst}},
\end{equation}
where $\Delta M_{\mathrm{star,burst}}$ is the total stellar mass formed during the starburst.
We set $f_{\mathrm{BH}}= 0.005$ to match the observed correlation between masses of host bulges and SMBHs at $z=0$.
Our model also reproduces the observationally estimated SMBH mass function.
We assume that the time evolution of the accretion rate is as follows:
\begin{equation}
\dot{M}_{\mathrm{BH}} (t) = \frac{\Delta M_{\mathrm{BH}}}{t_{\mathrm{life}}}\exp(-t/t_{\mathrm{life}}),
\end{equation}
where $t_{\mathrm{life}}$ is the quasar lifetime,
and that $t_{\mathrm{life}}$ scales with the dynamical time-scale of the host DM halo, that is, $t_{\mathrm{life}} \propto t_{\mathrm{dyn}}$.
We note that we allow for super-Eddington accretion in this model.
The cold gas accretion leads to quasar activity.
We assume that a fraction of the rest mass energy of the accreted gas is radiated.
The radiative efficiency is determined by the models we describe below.
Further details of our model of galaxy and quasar formation are given in \citet{2014ApJ...794...69E} and \citet{2016PASJ...68...25M}.

We also consider the radio-mode gas accretion and the feedback process.
In this mode the accreted gas powers radio jet that puts energy into the hot halo gas and prevents the hot gas from cooling and resultant star formation.
The radio-mode feedback occurs when (i) the timescale of gas cooling is sufficiently long compared with the dynamical timescale of the halo, and (ii) the cooling luminosity of the gas, which is assumed to balance the heating luminosity by the AGN, is sufficiently low compared with the Eddington luminosity.
Although this gas accretion grows the mass of SMBHs, this is not significant contribution for the entire mass growth of SMBHs.
This is because the mass growth of SMBHs is dominated by the gas accretion during major mergers (\citealt{2016PASJ...68...25M}).
Further details of the implementation of the radio-mode feedback are given in \citet{2016PASJ...68...25M}.

To examine the effect of the photon trapping in the accretion flow on the quasar host halo mass and the clustering, we adopt a simple model for the radiative efficiency $\eta \equiv L / \dot{M} c^2$ of a quasar, which imitates the effect of photon trapping for a high mass accretion rate,
\begin{equation}
\eta = \left( \frac{1}{\epsilon_{\mathrm{agn}}} + \frac{\dot{m}}{f_{\mathrm{Edd}}} \right)^{-1},
\end{equation}
where $\dot{m}$ is the normalised mass accretion rate $\dot{m}\equiv\dot{M}_{\mathrm{BH}}/\dot{M}_{\mathrm{Edd}}$ and $\epsilon_{\mathrm{agn}}$ is a free parameter.
The form of the model is the same as \citet*{2016MNRAS.461.4496S}.
In this model, the radiative efficiency decreases with increasing accretion rate, while it is constant for low accretion rates ($\dot{m}\lesssim10$).
The parameter, $\epsilon_{\mathrm{agn}}$, provides a constant radiative efficiency at such low accretion rates.
$f_\mathrm{Edd}$ corresponds to the maximum Eddington ratio for high mass accretion rates.
In our previous work, O16, $f_\mathrm{Edd}$ corresponed to $\infty$, which leads to
$\epsilon_{\mathrm{agn}} = \eta$.
We set the parameter $f_\mathrm{Edd} = $ 1, 2, and 10 to examine the effects of the maximum Eddington ratio on the quasar clustering.
We call the models with these parameters $\eta_{1}$, $\eta_{2}$, and $\eta_{10}$ model, respectively.
For comparison, we also adopt $f_\mathrm{Edd} = \infty$ as a model in which we allow for any super-Eddington luminosity ($\eta_{\infty}$ model).
We set $t_{\mathrm{life}} (z=0)$ and $\epsilon_{\mathrm{agn}}$ to match the luminosity function of quasars with the observation at $z=2$ in $\eta_{\infty}$ model.
We obtain $t_{\mathrm{life}} (z=0) = 2 \times 10^{7}$ yr and $\epsilon_{\mathrm{agn}} = 0.03$.
In O16, we employ this model.
To examine the photon trapping effect, we fix all parameters except for $f_\mathrm{Edd}$ and examine the effect of the change of $f_\mathrm{Edd}$ on the results.
We show the Eddington ratio in the models used here as a function of the accretion rate in Fig. \ref{fig:slimdisk}.
To obtain the quasar $B$-band luminosity, we use the bolometric correction from \citet{2004MNRAS.351..169M}.

\section{Results}
\label{sec:result}

\subsection{Accretion rate distribution}

\begin{figure}
	 \centering
	\includegraphics[width=0.9\columnwidth]{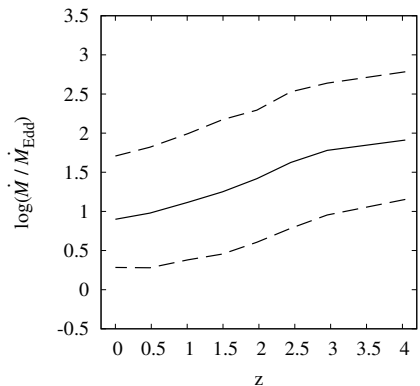}
    \caption{Evolution of normalised accretion rates for quasars.
    The solid line denotes the median.
    The dashed lines show the 10th and 90the percentiles.
    Note that we use equation (\ref{eq:edd}) for the definition of $\dot{M}_{\mathrm{Edd}}$.
    The Eddington ratio $L/L_{\mathrm{Edd}} = \eta \dot{M}_{\mathrm{BH}}/\dot{M}_{\mathrm{Edd}}$, where $\eta \leq 0.03$ in our model.
    Thus, the luminosity of most quasars is sub-Eddington at $z\sim0$.
    }
    \label{fig:z_edd_ratio}
\end{figure}

To show the importance of the model of the radiative efficiency during super-Eddington accretion, in particular, at high redshift, we first show the redshift evolution of the normalised accretion rate for our model quasars in Fig. \ref{fig:z_edd_ratio}.
This figure shows that the normalised accretion rate increases with redshift.
The super-Eddington accretion is expected to occur, in particular, at high redshift.
We expect that the slim accretion disc model affects the quasar clustering more strongly at higher redshift.

\subsection{Quasar host halo mass}
\label{subsec:halo_mass}

\begin{figure*}
  \begin{center}
    \includegraphics[width=160mm]{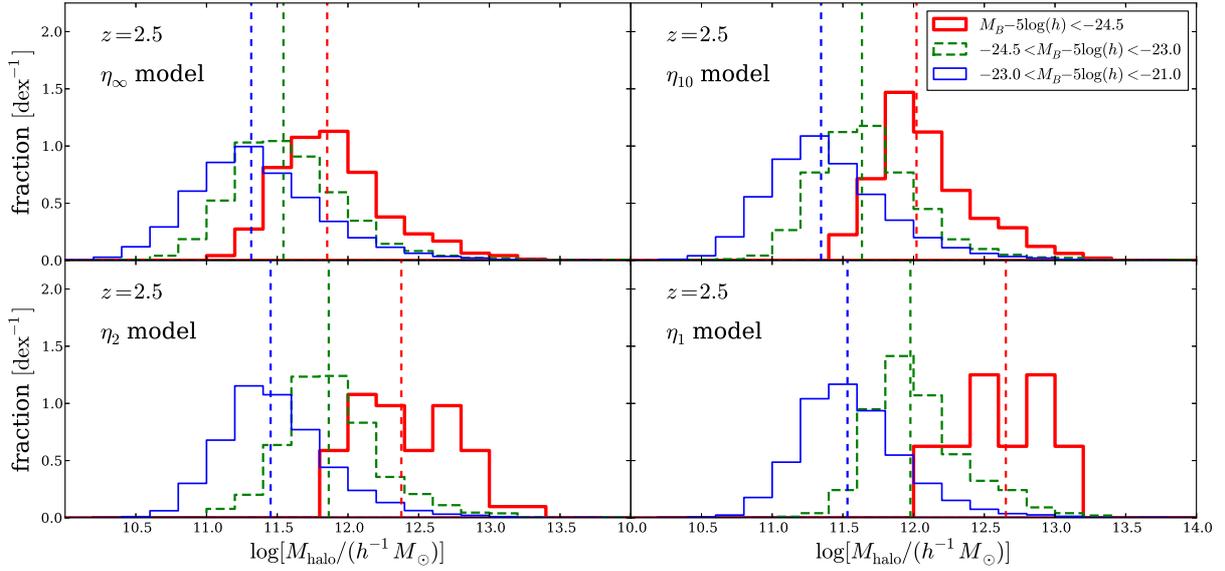}
    \caption{Mass distributions of DM haloes hosting bright (red, thick solid), intermeiate (green, dashed) and faint (blue, thin solid) quasars at $z=2.5$.
    Each panel corresponds to a different model of radiative efficiency, as indicated by the legend.
    The vertical dashed lines are the median masses of the distributions.}
\label{fig:m_dist}
  \end{center}
\end{figure*}

Here, we investigate the impact of the model of radiative efficiency on the quasar host halo mass.
Fig.~\ref{fig:m_dist} shows the mass distributions of the quasar host haloes for three quasar-magnitude bins and their median halo masses (dashed lines) at $z=2.5$ when the number density of quasars has a noticeable peak.
We define bright, intermediate, and faint quasars as those with $B$-band magnitudes $M_{B} - 5\log h < -24.5$, $-24.5 < M_{B} - 5\log h < -23.0$ and $-23.0 < M_{B} - 5\log h < -21.0$, respectively.
We show different models of radiative efficiency in different panels.

Fig.~\ref{fig:m_dist} clearly shows that the radiative efficiency, in other words, the maximum value of the quasar luminosity affects the quasar host halo mass distribution.
We see that decreasing the upper limit of the quasar luminosity, the median host halo mass increases.
For $\eta_{\infty}$ model, there is only a weak dependence of median halo mass on quasar magnitude.
On the other hand, for $\eta_{1}$ model, there is a strong dependence of the median halo mass on quasar magnitude.
While, for $\eta_{\infty}$ model, the difference between the median halo masses of bright and faint quasars is $\sim$~0.5~dex, for $\eta_{1}$ model, the difference reaches $\sim$~1.2~dex.
This trend can be understood as follows.
Because of the fact that more massive haloes have more massive BHs in our model, when two quasars in haloes with different masses have same luminosity, the quasar in the less massive halo needs to have a high Eddington ratio.
As a result, decreasing the upper limit of the quasar luminosity, the median host halo mass increases.
This trend appears in the quasar bias described in Sec. \ref{subsec:bias}.

In all magnitude range, the median host halo mass increases with the decreasing upper limit of quasar luminosity, although the upper limit affects the median stronger for bright quasars than that of faint quasars.
The median host halo mass of bright quasars increases more than that of faint quasars by adopting the upper limit of quasar luminosity.
This is because more luminous quasars within less massive haloes have higher Eddington ratios.
Therefore, in this case, the median host halo mass depends on the quasar luminosity.

\begin{figure*}
  \begin{center}
    \includegraphics[width=130mm]{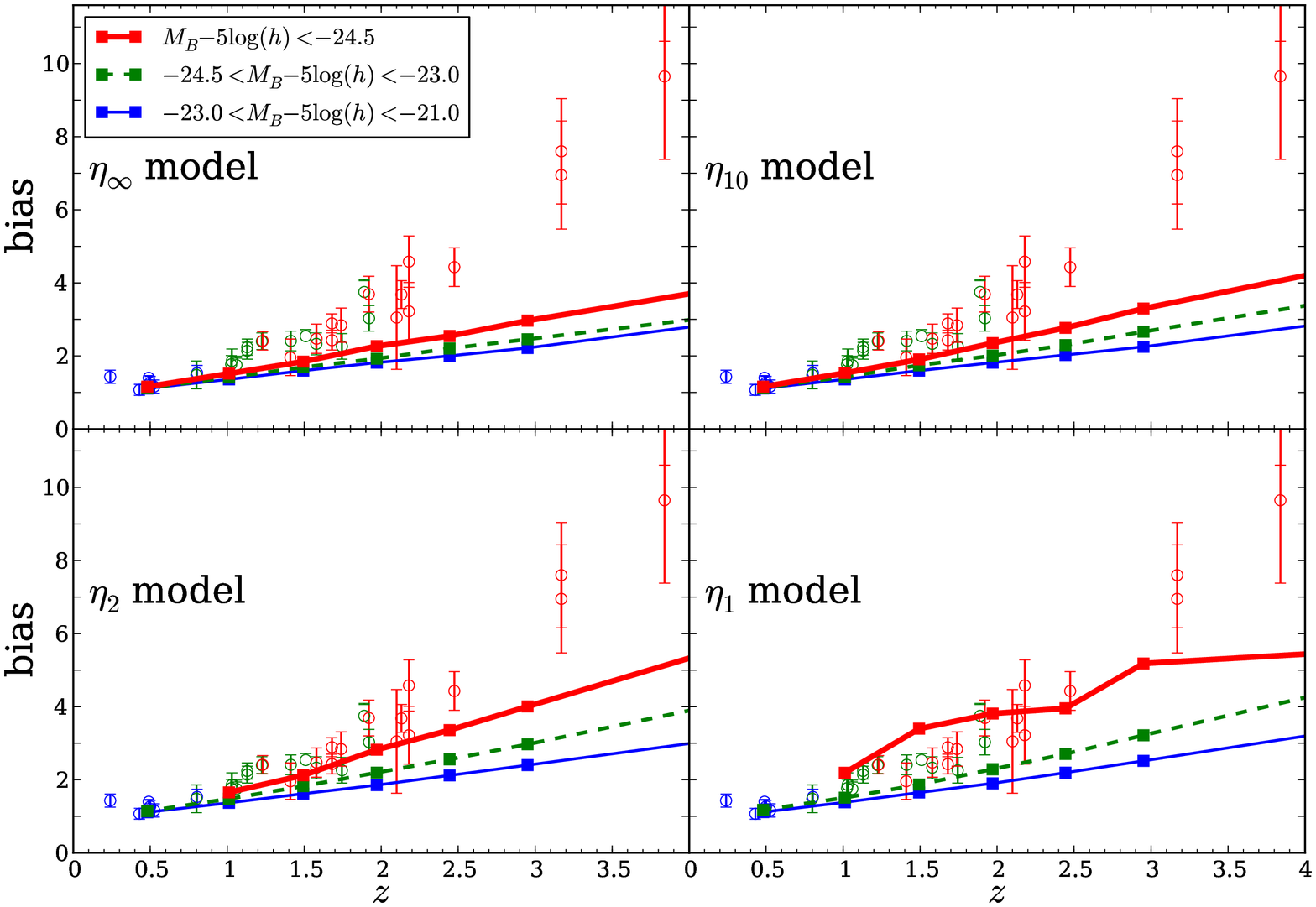}
    \caption{Redshift evolution of the bias of bright (red, thick solid), intermediate (green, dashed) and faint (blue, thin solid) quasars.
    Each panel corresponds to a different model of radiative efficiency, as indicated by the legend.
    Observational results by the large-scale surveys of SDSS and 2QZ \citep[][]{2004MNRAS.355.1010P, 2005MNRAS.356..415C, 2009MNRAS.397.1862P, 2009ApJ...697.1634R, 2009ApJ...697.1656S} are also plotted (small filled circles and error bars), whose magnitudes are converted to $M_{B}$.
    The color coding represents the luminosity of quasars in the same way as for the models.}
\label{fig:bias}
  \end{center}
\end{figure*}

\subsection{Quasar bias}
\label{subsec:bias}

We estimate the quasar bias using the median host halo mass and the following equation derived by \citet*{2001MNRAS.323....1S}.
This is because the bias of quasars is primarily determined by the median host halo mass.
\citet{2001MNRAS.323....1S} relates the halo bias and its mass:
\begin{align}
b (M,z) &= 1 + \frac{1}{\sqrt{a} \delta_c} \biggl[ \sqrt{a} (a \nu^2) + \sqrt{a}b(a\nu^2)^{1-c} \biggr. \nonumber \\
& \biggl. \quad - \frac{(a\nu^2)^c}{(a\nu^2)^c + b(1-c)(1-c/2)} \biggr],
\end{align}
where $a = 0.707$, $b = 0.5$, $c = 0.6$, $\delta_c$ = 1.686 is the critical overdensity required for collapse and $\nu = \frac{\delta_c}{\sigma (M) D(z)}$.
$D(z)$ is the linear growth factor and $\sigma (M)$ is the variance of mass fluctuations.
We have confirmed that this method works well in O16.
The redshift evolution of the bias is shown in Fig. \ref{fig:bias}, where we show different models of radiative efficiency in different panels.
Decreasing the upper limit of the quasar luminosity, the quasar bias increases.
The upper limit of the quasar luminosity affects the bias for brighter quasars and at higher redshift more strongly.
At $z\gtrsim2$, the bias of bright quasars increases significantly for $\eta_{1}$ and $\eta_{2}$ models.
This is because of the following two reasons.
First, in these cases only massive BHs become bright quasars due to the upper limit $L_{\mathrm{max}}$, which is proportional to $M_{\mathrm{BH}}$, i.e. $L_{\mathrm{max}} \sim L_{\mathrm{Edd}} \propto M_{\mathrm{BH}}$.
Second, in our model more massive BHs reside in more massive haloes, which is in agreement with the observations (e.g. \citealt{2002ApJ...578...90F, 2006MNRAS.373..613F}, but see also \citealt{2015ApJ...803....5S}).
Therefore, luminous quasars having massive BHs show the strong clustering in these cases.
On the other hand, the quasar bias hardly changes for all magnitude bins at low redshift ($z\sim1$) even for $\eta_{1}$ and $\eta_{2}$ models.
This is because most quasars have luminosities below the Eddington luminosity in this epoch.
These quasars are not influenced by the upper limit.

We compare our results with observations.
We compile observational results by the large-scale surveys of SDSS and 2QZ (\citealt*{2004MNRAS.355.1010P}; \citealt{2005MNRAS.356..415C}; \citealt{2009MNRAS.397.1862P}; \citealt{2009ApJ...697.1634R}; \citealt{2009ApJ...697.1656S}).
For $\eta_{1}$ model, the bias of bright quasars is in agreement with the general increasing trend of the bias shown in observational data.
In addition, for $\eta_{2}$ model, the bias is in agreement with the observation at $1.5\lesssim z \lesssim 2$.
Future surveys of quasars will allow us to make more accurate comparison to the model.

\section{Summary and Discussion}
\label{sec:discussion}

In this Letter, we have investigated the large-scale clustering of quasars, including the models of radiative efficiency during super-Eddington accretion using our semi-analytic model, $\nu^2$GC.
We have shown that the model of radiative efficiency strongly affects the quasar clustering.
While the quasar bias has no significant dependence on quasar luminosity for the no-limit model, it has significant dependence on quasar luminosity for the model in which quasar luminosity is limited by its Eddington luminosity.
This is because only massive BHs become high luminosity quasars due to the limit, and, in general, such massive SMBHs reside in massive haloes.

The luminosity dependence of the quasar bias is a reflection of the super-Eddington accretion model.
This indicates that quasar clustering studies provide a constraint on the accretion disc model.
The existence of the super-Eddington accretion discs are observationally suggested. 
\citet{2014MNRAS.438..672N} have estimated the fraction of super-Eddington accretion discs in AGNs with different BH masses and accretion rates and have suggested that the super-Eddington accretion discs are very common among AGNs at various BH masses and redshift ranges.
Super-Eddington accretion may also leave an imprint on the spectra of the radiation 
from the accreted gas on to SMBHs.
\citet{2015MNRAS.454.3771P} have simulated the spectra by using radiation-hydrodynamic and spectral synthesis codes, and have shown the
characteristics of the spectra for Eddington-limited accretion and super-Eddington accretion flows.
Studying the spectra from high-\textit{z} AGNs would help in understanding the importance of super-Eddington
accretion on to high-\textit{z} SMBHs.
The quasar clustering at high-\textit{z} we have investigated in this Letter may give a complementary method to investigate the AGNs with the super-Eddington accretion discs.

Our results show that models should take into account the photon trapping effect when 
considering the super-Eddington accretion.
There are many studies investigating the quasar clustering using semi-analytic models
(e.g. \citealt{2002MNRAS.332..529K}; \citealt*{2003PASJ...55..133E}; \citealt{2009MNRAS.396..423B}; \citealt{2013MNRAS.436..315F}; \citealt{2016MNRAS.456.1073G}) or cosmological hydrodynamical simulations \citep*[e.g.][]{2011MNRAS.413.1383D, 2017MNRAS.466.3331D}, however, most of these studies do not allow for the super-Eddington mass accretion.
Recent theoretical (\citealt{2015MNRAS.452.1922P}) and observational (\citealt{2014MNRAS.438..672N}) studies suggest that models including the super-Eddington accretion and the resultant photon trapping effect could help us understand further physical mechanisms which cause the observed quasar clustering.

We have shown that for $\eta_{1}$ model, the bias of bright quasars is in agreement with the observation at $z\gtrsim2$.
Previous studies (e.g. O16; \citealt{2016ApJ...832...70A}) claim that it is difficult to explain the strong quasar clustering at high redshift ($z>2$) \citep{2007AJ....133.2222S, 2009ApJ...697.1656S, 2016ApJ...832...70A}.
Our results support a picture in which BHs grow super-Eddington accretion, and could solve the problem of the strong clustering.
Future observations of high-redshift quasar clustering will allow us to make more accurate comparisons to the model we introduced here and to estimate the role of super-Eddington accretion process during BH growth.
For X-ray selected AGNs, \citet{2010ApJ...716L.209C} have shown the weak X-ray luminosity dependence of AGN clustering.
Future theoretical studies are needed to explain the luminosity dependent clustering of these relatively low luminosity AGNs.

\section*{Acknowledgements}

We are grateful to the referee for helpful suggestions which improve the Letter.
We thank K. Ohsuga, M. Akiyama, H. Ikeda, T. Nagao and K. Wada for useful comments and discussion.
This research was supported by World Premier International Research Center Initiative (WPI), MEXT, Japan and by
MEXT as ``Priority Issue on Post-K computer'' (Elucidation of the Fundamental Laws and Evolution of the Universe) and JICFuS.
This study has been funded by MEXT/JSPS KAKENHI Grant Number 25287041, 15K12031 and by Yamada Science Foundation, MEXT HPCI STRATEGIC PROGRAM.
RM acknowledge support from JSPS KAKENHI Grant Numbers JP 15H05896.
TO has been supported by MEXT KAKENHI Grant 16H01085.
MN has been supported by MEXT KAKENHI Grant 17H02867.



\bibliographystyle{mnras}
\bibliography{ref} 








\bsp	
\label{lastpage}
\end{document}